# Investigation of shape coexistence in $^{118-128}Te$ isotopes

H. Sabri[a1], Z. Jahangiri[b], M. A. Mohammadi[a]

[a] Department of Physics, University of Tabriz, Tabriz 51664, Iran.

[b] Physics Department, Payame Noor University, Tehran 19395-4697, Iran.

[1] Corresponding Author E-mail: h-sabri@tabrizu.ac.ir

## Abstract

In this paper, we have considered the interplay between phase transitions and configuration mixing of intruder excitations in the $^{118\text{-}128}Te$ isotopes. A transitional interacting boson model Hamiltonian in both IBM-1 and IBM-2 versions which are based on affine $SU(1,1)$ Lie Algebra are employed to describe the evolution from spherical to deformed gamma unstable shapes along the chain of *Te* isotopes. The excitation energies, *B*(*E0*) and *B*(*E2*) transition rates are rather well reproduced in comparison with experimental counterparts when the weight of *SO*(*6*) limit is increased in Hamiltonian. Also our results show obvious relations between the configuration mixing ratio and quadrupole, hexadecapole and hexacontatetrapole deformation values in this isotopic chain.

## 1. Introduction

Shape coexistence has been observed in many mass regions throughout the nuclear chart and has become a very useful paradigm to explain the competition between the monopole part of the nuclear effective force that tends to stabilize the nucleus into a spherical shape, in and near to shell closures, and the strong correlations (pairing, quadrupole in particular) that favors the nucleus into a deformed shapes in around mid-shell regions [1-15]. The even-mass tellurium isotopes are part of this interesting region beyond the closed proton shell where the level structure has resisted detailed theoretical understanding. Although these nuclei were thought to be vibrational, the comparable strengths of the $2_2^+ \to 2_1^+$ and the $2_1^+ \to 0_1^+$ transitions in many isotopes and the systematic variation of the $0^+$ level energy provides a clue suggesting that they might have a transitional character. Moreover, most of the measured quadrupole moments of the $2_1^+$ states are non-zero and some are quite large with values that are comparable in magnitude with those of rotational nuclei. The early study of the collective potential-energy surface and nuclear structure of $^{124}Te$ excluded a rotational behavior since the level spectrum does not obey the $l(l+1)$ rule. A vibrational character was also excluded for these nuclei by the absence of a $0^+$ level in the energy region of the two-phonon triplet. The $\gamma-$independence of the potential-energy surface and level structure of some tellurium, xenon and barium isotopes were predicted at the same time and these features have been supported by several experimental and theoretical studies [4-25].

From a theoretical side, present-day methods starting from both the nuclear shell model, or approaching the atomic nucleus using mean-field methods have resulted in developments of both new algorithms as well as making use of the increased computing possibilities [26-28]. Although, it is very difficult to treat the structure of such nuclei in terms of conventional mean field theories since they are neither vibrational nor rotational. The present status has evolved in a situation where the conditions for shape coexistence to occur are becoming understood. Recent large-scale shell-model studies (LSSM), using diagonalization in a very large many-open shell basis in various mass regions [29] or making use of an advanced Monte-Carlo shell-model approach (MCSM) [28], have been carried out. Besides, the concept to start from deformed average potentials and calculating the total energy curves as a function of deformation has been explored [30-32]. However, recent studies [33], using a microscopic approach to determine the optimal mean fields, even going beyond when bringing in the nuclear dynamics, have given quantitative results - using both Skyrme forces [33-36] and Gogny forces [37-41], as well as making use of a relativistic mean-field approach [42-47] - that are indicative of the above mechanism. Moreover, attempts have been made and are still improved to extract a Bohr Hamiltonian [48-49] starting from a microscopic basis [50-52]. From a microscopic shell-model point of view, the hope to treat on equal footing the large open neutron shell from *N*=82 down to and beyond the mid-shell *N*=66 region, jointly with the valence protons in the *Cd*, *Sn*, *Te* and *Xe* nuclei, even including proton multi-particle multi-hole (*mp-nh*) excitations across the *Z*=50 shell closure, is beyond present computational possibilities. The truncation of the model space, however, by concentrating on nucleon pair modes (mainly $0^+$ and $2^+$ coupled pairs, to be treated as bosons within the interacting boson approximation (IBM) [53]), has made calculations feasible, even including pair excitations across the *Z*=50 shell closure [54-55].

In the Interacting Boson Model (IBM) [53-60], which describes the nuclear structure of even–even nuclei within the $U(6)$ symmetry, possessing $U(5)$, $SU(3)$ and $O(6)$ dynamical symmetry limits, shape phase transitions have been studied 25 years ago with using the classical limit of the model. These descriptions point out that there is a second order shape phase transition between $U(5)$ and $O(6)$ limits, namely between spherical and deformed gamma unstable limits which *Te* isotopes are expected to lie in this transitional region. The analytic description of nuclear structure at the critical point of phase transitions has attracted extensive interest in the recent decades. One has to employ some complicated numerical methods to diagonalize the transitional Hamiltonian in these situations but Pan *et al* in Refs.[61-62] have been proposed a new solution which was based on affine $SU(1,1)$ algebraic technique and explores the properties of nuclei have classified in the $U(5) \leftrightarrow SO(6)$ transitional region of IBM.

It was long believed, the Tellerium isotopes were good examples of the quadrupole vibrational nuclei, namely $U(5)$ nuclei [10-19]. However, during the last few years, new experimental data and calculations have led to a modified picture on these nuclei. These mean *Te* isotopes appear to evolve from the $U(5)$ to $O(6)$ -like structure in IBM classification.

In this study, we have determined the energy levels, $B(E0)$ and $B(E2)$ transition probabilities of $^{118\text{-}128}Te$ isotopes by using a transitional Hamiltonian which is based on the affine $SU(1,1)$ algebraic technique. We also considered the variation of control parameter in related to quadrupole, hexadecapole and hexacontatetrapole deformation values in the isotopic chain. Our results propose a spherical to $\gamma-$soft shape transitions in these nuclei and also explore the intruder states.

This paper is organized as follows: section 2 briefly summarizes theoretical aspects of transitional Hamiltonian and affine $SU(1,1)$ algebraic technique. Section 3 includes the results of this study about control parameters, transition probabilities for this isotopic chain and section 4 is devoted to summarize and some conclusion.

## 2. Theoretical description of transitional region and shape coexistence

As have been shown in various spectroscopic selective experiments, *e.g.* transfer reactions in particular, very near to closed shells (the In and *Sb* nuclei at *Z*=50 but also in other mass regions, *e.g.* the *Tl* and *Bi* nuclei at *Z*=82) some low-lying extra states, so-called intruder states, have been observed with a conspicuous energy dependence on the number of free valence neutrons, hinting for 2*p*-2*h* excitations as their origin [34-35]. If these excitations are proton excitations combined with the neutron degree of freedom appearing on both sides of the *Z*=50 closed shell, such as condition which are available for *Te* isotopes, it is a natural step to suggest that low-lying extra $0^+$ excitations will also show up in the even-even nuclei in between. Because the *Te* isotopes with a large number of valence neutrons are situated near to the $\gamma$ -stability line, they could be studied [8].

Since making a 1*p*-1*h* excitation across the closed shell at Z=50 takes about 4.5 *MeV* (the proton shell gap), the unperturbed energy for 2p-2h excitations comes up to about 9 *MeV*. Even though pairing amongst the particles and holes will lower the energy in an important way to 4-5 *MeV*, this is still far from the observed excitation energy of 1.7 *MeV*. Some essential element is missing when starting from the spherical intrinsic symmetry of the shell model.

One way to come around is breaking the spherical symmetry and allowing the mean field to acquire quadrupole deformation thereby giving rise to the possibility that spherical orbits split and the large spherical shell gap at *Z*=50 and also at *Z*=82 rapidly vanishes. Calculations have been carried out over the years using deformed mean-field studies, *e.g.* Nilsson model, deformed Woods-Saxon, Hartree-Fock-Bogoliubov studies) and we would like to cite in particular [37-30]. The deformed field essentially points out to need for the quadrupole component in the mean field as the agent for the increased binding energy. Knowing this, and having experimental knowledge of the fact that 1*p*-1*h* (in odd-mass nuclei) and 2*p*-2*h* (in even-even nuclei) are present in these states, it is tempting to incorporate this in a spherical shell-model description. By invoking a schematic model that was discussed in detail in [34] it is possible to evaluate the excitation energy of a 2*p*-2*h* configuration

$$E_{\mathrm{int}r.}\left(0^+\right) = \left\langle 0_I^+ \left| H \right| 0_I^+ \right\rangle - \left\langle 0_{GS}^+ \left| H \right| 0_{GS}^+ \right\rangle \quad ,$$

in which the index *I* denotes the nucleon distribution in the intruder state and *GS* the distribution in the ground state. Using a pair distribution for the neutrons, combined with a 2*p*-2*h* excitation and a 0*p*-0*h* excitation for the intruder and regular state, respectively, one can derive the expression

$$E_{intr.}(2p-2h) \cong 2\left(\varepsilon_p - \varepsilon_h\right) + \Delta E_M\left(2p-2h\right) - \Delta E_{pair} + \Delta E_Q\left(2p-2h\right) \quad ,$$

where the various terms describe the unperturbed energy to create the 2*p*-2*h* configuration, a monopole correction due to a change in proton single-particle energy while changing the neutron number, the pairing-energy correction because $0^+$- coupled pairs are formed, and the quadrupole binding energy originating from the proton-neutron force, respectively.

In calculating the neutron number dependence of the 2*p*-2*h* intruder $0^+$ configurations we have to determine the quadrupole energy contribution and this we do by using the *SU*(3) expression given in [34], i.e.,

$$\Delta E_Q \cong 2\kappa \Delta N_\pi N_\nu \quad ,$$

in which $\Delta N_\pi$ denotes the number of pairs excited out of the closed shell configuration at *Z*=50, i.e., $\Delta N_\pi = 2$ for a 2*p*-2*h* excitation.

This approach points out that the essential elements are the strong pairing interactions amongst the particles and the holes that make up for the excited configuration and the strong quadrupole proton-neutron forces. It is precisely here that early contacts between the disconnected ``spaces" of interacting boson within a valence space only and the *p-h* excitations of the core itself showed up. In a lowest order approximation, one can think of the 2*p* and the 2*h* parts to bring in two extra bosons increasing the active model space from *N* to *N+2* bosons and carry out separate calculations for both spaces introducing a coupling between them by using a mixing Hamiltonian

$$H_{mix} = \alpha \left( s^\dagger s^\dagger \right)^0 + \beta \left( d^\dagger d^\dagger \right)^0 + h.c.$$

The presence of these extra states, characterized by 2*p*-2*h* excitations across the *Z*=50 shell closure, has become a fingerprint especially near the *N*=66 mid-shell region. Moreover, the interference between the regular vibrational states and these intruder states that contain a much larger collectivity, shows up as drastic modifications of the regular vibrational *E*2 intensity ratios.

A full shell-model study of the *Te* nuclei, with neutrons moving all through the full valence space of *N*=50 towards *N*=82, at the same time incorporating, besides the two proton holes outside of *Z*=50, the 2*p*-2*h* excitations that show up in the mid-shell neutron region (around *N*~66), is out of reach. Only when approaching the neutron shell closure at *N*=50 i.e. *N*=52,54 and for the heavy nuclei near *N*=82 considering the cases with *N*=80,78 and also beyond, at *N*=84, full shell-model studies can become feasible. Therefore, the study of these extreme heavy nuclei is important since it may shed light on the way how collective quadrupole states (with anharmonicities included) may go over into the shell-model structure: there should be some region of overlap which can give us very interesting information.

As mentioned in Refs.[38-34], the nuclear shell model is not in a position to be used for a reliable computation of the low-energy properties of the full range of *Te* isotopes. This means that one has to resort to a suitable truncation of the shell model, such a model has been developed in Ref.[36] which is based on quasiparticle random-phase approximation (QRPA) or use algebraic approach to explore the considered nuclei. On the other hand, the drawback of those calculations is that one easily gets involved with a lot of parameters and unless one has some physics guidance the detailed agreement needs some caution. To aim these problems, we have used an affine $SU(1,1)$ Lie Algebra [61-62] to construct the transitional Hamiltonian. Although the results of this approach are somewhat different from those of the IBM, but as have presented in Refs.[61-64], a clear correspondence with the description of the geometrical model is obvious for this transitional region.

## 2.1. An affine $SU(1,1)$ algebraic solution for $U(5) \leftrightarrow SO(6)$ Transitional Hamiltonian

The $SU(1,1)$ Algebra has been described in detail in Refs.[61-62]. Here, we briefly outline the basic ansatz and summarize the results. The Lie algebra which corresponds to the $SU(1,1)$ group is generated by $S^{\nu}, \nu = 0$ and $\pm$, which satisfies the following commutation relations

$$[S^0, S^{\pm}] = \pm S^{\pm} \qquad , \qquad [S^+, S^-] = -2S^0 \tag{1}$$

The Casimir operator of $SU(1,1)$ group can be written as

$$\hat{C}_2 = S^0(S^0 - 1) - S^+S^- \qquad , \tag{2}$$

Representations of $SU(1,1)$ are determined by a single number $\kappa$, thus the representation of Hilbert space is spanned by orthonormal basis $|\kappa\mu\rangle$ where $\kappa$ can be any positive number and $\mu = \kappa, \kappa+1, \ldots$. Therefore,

$$\hat{C}_2(SU(1,1))|\kappa\mu\rangle = \kappa(\kappa - 1)|\kappa\mu\rangle \qquad , \qquad S^0|\kappa\mu\rangle = \mu|\kappa\mu\rangle \tag{3}$$

In IBM, the generators of $d-$boson pairing algebra is created by

$$S^+(d) = \frac{1}{2}(d^{\dagger}.\, d^{\dagger}) \quad , \quad S^-(d) = \frac{1}{2}(\tilde{d}.\, \tilde{d}) \quad , \quad S^0(d) = \frac{1}{4}\sum_{\nu}(d_{\nu}^{\dagger}d_{\nu} + d_{\nu}d_{\nu}^{\dagger}) \tag{4}$$

Similarly, $s-$boson pairing algebra forms another $SU^s(1,1)$ algebra which is generated by

$$S^+(s) = \frac{1}{2}s^{\dagger 2} \quad , \quad S^-(s) = \frac{1}{2}s^2 \quad , \quad S^0(s) = \frac{1}{4}(s^{\dagger}s + ss^{\dagger}) \tag{5}$$

On the other hand, the infinite dimensional $SU(1,1)$ algebra is generated by using of [61-64]

$$S_n^{\pm} = c_s^{2n+1} S^{\pm}(s) + c_d^{2n+1} S^{\pm}(d) \quad , \quad S_n^0 = c_s^{2n} S^0(s) + c_d^{2n} S^0(d) \tag{6}$$

Where $c_s$ and $c_d$ are real parameters and $n$ can be $0, \pm 1, \pm 2, \ldots$. These generators satisfy the commutation relations,

$$[S_m^0, S_n^{\pm}] = \pm S_{m+n}^{\pm} \quad , \quad [S_m^+, S_n^-] = -2 S_{m+n+1}^0 \tag{7}$$

Then, $\{S_m^{\mu}, \mu = 0, +, -; \pm 1, \pm 2, \ldots\}$ generates an affine Lie algebra $SU(1,1)$ without central extension. By employing the generators of $SU(1,1)$ Algebra, the following Hamiltonian is constructed for the transitional region between $U(5) \leftrightarrow SO(6)$ limits [61-62]

$$\hat{H} = g\ S_0^+ S_0^- + \varepsilon\ S_1^0 + \gamma\ \hat{C}_2(SO(5)) + \delta\ \hat{C}_2(SO(3)) \tag{8}$$

$g, \varepsilon, \gamma$ and $\delta$ are real parameters where $\hat{C}_2(SO(3))$ and $\hat{C}_2(SO(5))$ denote the Casimir operators of these groups. It can be seen that Hamiltonian (8) would be equivalent with $SO(6)$ Hamiltonian if $c_s = c_d$ and with $U(5)$ Hamiltonian when $c_s = 0$ & $c_d \neq 0$. Therefore, the $c_s \neq c_d \neq 0$ requirement just corresponds to the $U(5) \leftrightarrow SO(6)$ transitional region. In our calculation we take $c_d$ (=1) constant value and $c_s$ vary between 0 and $c_d$.

Eigenstates of Hamiltonian (8) can obtain with using the Fourier-Laurent expansion of eigenstates and $SU(1,1)$ generators in terms of unknown $c-$number parameters $x_i$ with $i = 1, 2, \ldots, k$. It means, one can consider the eigenstates as [61-62]

$$\left| k; \nu_s \nu n_{\Delta} LM \right\rangle = \sum_{n_i \in Z} a_{n_1} a_{n_2} \ldots a_{n_k} x_1^{n_1} x_2^{n_2} \ldots x_k^{n_k} S_{n_1}^+ S_{n_2}^+ \ldots S_{n_k}^+ \left| lw \right\rangle \quad , \tag{9}$$

Due to the analytical behavior of wavefunctions, it suffices to consider $x_i$ near zero. With using the commutation relations between the generators of $SU(1,1)$ Algebra, i.e. Eq.(7), wavefunctions can be considered as:

$$\left| k; \nu_s \nu n_{\Delta} LM \right\rangle = N S_{x_1}^+ S_{x_2}^+ \ldots S_{x_k}^+ \left| lw \right\rangle \quad , \tag{10}$$

where $N$ is the normalization factor and

$$S^{+}_{x_i} = \frac{c_s}{1-c_s^2 x_i} S^{+}(s) + \frac{c_d}{1-c_d^2 x_i} S^{+}(d) \qquad , \qquad (11)$$

The c-numbers $x_i$ are determined through the following set of equations

$$\frac{\epsilon}{x_i} = \frac{g{c_s}^2(\nu_s + \frac{1}{2})}{1-{c_s}^2 x_i} + \frac{gc_d^2(\nu + \frac{5}{2})}{1-c_d^2 x_i} - \sum_{i \neq j} \frac{2}{x_i - x_j} \qquad \text{for i=1,2,...,k} \qquad (12)$$

Eigenvalues of Hamiltonian (8), i.e. $E^{(k)}$, can be expressed as [61-62]

$$E^{(k)} = h^{(k)} + \gamma\nu(\nu+3) + \delta L(L+1) + \varepsilon\Lambda_1^0 \qquad , \qquad \Lambda_1^0 = \frac{1}{2}[c_s^2(\nu_s + \frac{1}{2}) + c_d^2(\nu + \frac{5}{2})] \qquad (13)$$

Which

$$h^{(k)} = \sum_{i=1}^{k} \frac{\varepsilon}{x_i} \quad , \qquad (14)$$

The quantum number $k$, is related to total boson number $N$, by

$$N = 2k + \nu_s + \nu$$

To obtain the numerical results for $E^{(k)}$, we have followed the prescriptions have introduced in Refs.[61-62], namely a set of non-linear Bethe-Ansatz equations (BAE) with $k$ – unknowns for $k$ – pair excitations must be solved. To this aim we have changed the variables as

$$\epsilon = \frac{\varepsilon}{g} (g = 1 \ kev \ [61\text{-}62]) \qquad \text{c} = \frac{c_s}{c_d} \leq 1 \qquad \text{y}_i = c_d^2 x_i$$

so, the new form of Eq.(12) would be

$$\frac{\epsilon}{y_i} = \frac{c^2(\nu_s + \frac{1}{2})}{1-c^2 y_i} + \frac{(\nu + \frac{5}{2})}{1-y_i} - \sum_{i \neq j} \frac{2}{y_i - y_j} \qquad \text{for i=1,2,...,k} \qquad (15)$$

We have solved Eq. (15) with definite values of $c$ and $\varepsilon$ for $i = 1$ to determine the roots of Beth-Ansatz equations (BAE) with specified values of $\nu_s$ and $\nu$ , similar to procedure which have done in Refs.[61-62]. Then, we have used “Find root” in the Maple17 to get all $y_j^{'}$ s. We carry out this procedure with different values of $c$ and $\varepsilon$ to provide energy spectra (after inserting $\gamma$ and $\delta$ ) with minimum variation as compared to the experimental counterparts;

$$\sigma = (\frac{1}{N_{tot}} \sum_{i,\, tot} \left| E_{\exp}(i) - E_{cal}(i) \right|^{2})^{1/2}$$

which $N_{tot}$ is the number of energy levels where are included in extraction processes. We have extracted the best set of Hamiltonian's parameters, i.e. $\gamma$ and $\delta$ , via the available experimental data [65-72] for excitation energies of selected states, $0_1^+, 2_1^+, 4_1^+, 0_2^+, 2_2^+, 4_2^+$ and *etc*, e.g. 12 levels up to $2_4^+$ , or two neutron separation energies for nuclei which are considered in this study. In summary, we have extracted $\gamma$ and $\delta$ externally from empirical evidences and other quantities of Hamiltonian, e.g. $c$ and $\varepsilon$ would determine through the minimization of $\sigma$ .

The results of IBM-1 which present in the following, cannot reproduce energy spectra and *B*(E2) transition probabilities exactly in comparison with experimental counterparts. As have been explained in Refs.[61-64] extensively, in the Tellurium isotopes, the normal vibrational states are mixed with deformed intruder states and therefore one may use the IBM-2 calculation to explore three phonons or 2p-4h states with more accuracy in comparison with IBM-1.

To this aim we have used a same formalism to extend the IBM-2 calculation via $SU(1,1)$ lie algebra. The details of this approach have presented in Refs.[61-62] and we explore the final results here. In IBM-2 case, the Hamiltonian can be considered as

$$\hat{H} = g\, S_0^+ S_0^- + \varepsilon\, S_1^0 + \gamma_1\, \hat{C}_2(SO_\pi(5)) + \gamma_2\, \hat{C}_2(SO_\nu(5)) + \delta_1\, \hat{C}_2(SO_\pi(3)) + \delta_2\, \hat{C}_2(SO_\nu(3)) + \delta\, \hat{C}_2(SO(3)) \qquad (16)$$

In this Hamiltonian

$$S_n^{\pm} = \sum_t c_{s;\,t}^{2n+1} S^{\pm}(s;t) + c_{d;\,t}^{2n+1} S^{\pm}(d;t) \quad , \quad S_n^0 = \sum_t c_{s;\,t}^{2n} S^0(s;t) + c_{d;\,t}^{2n} S^0(d;t) \qquad (17)$$

And the sum is over the proton, $\pi$, and neutron, $\nu$ , indices. The eigenstates of Eq. (16) can be expressed as

$$\left| k;\beta;\nu_s^\pi,\nu_s^\nu,\nu^\pi,\nu^\nu;n_\Delta^\pi L_\pi, n_\Delta^\nu L_\nu; LM \right\rangle = N S_{x_1}^+ S_{x_2}^+ \ldots S_{x_k}^+ \left| lw \right\rangle \quad , \qquad (18)$$

where $2k = N_\pi + N_\nu - \nu_s^\pi - \nu_s^\nu - \nu^\pi - \nu^\nu$ and

$$S^{+}_{x_i} = \sum_t \frac{c_{s;\,t}}{1-c^2_{s;\,t}x_i} S^{+}(s;t) + \frac{c_{d;\,t}}{1-c^2_{d;\,t}x_i} S^{+}(d;t) \quad , \tag{19}$$

Similar to IBM-1 case, the *c*-number $x_i$ satisfy a set of equations similar to (12)

$$\frac{\epsilon}{x_i} = \sum_t g\left(\frac{c_{s;\,t}{}^2(\nu^t_s + \frac{1}{2})}{1-c_{s;\,t}{}^2 x_i} + \frac{c^2_{d;\,t}(\nu^t + \frac{5}{2})}{1-c^2_{d;\,t}x_i}\right) - \sum_{i\neq j} \frac{2}{x_i - x_j} \qquad \text{for i=1,2,...,k} \tag{20}$$

Finally, the eigenvalues of Eq.(16) can be expressed as

$$E^{(k)} = \sum_{i=1}^{k} \frac{\varepsilon}{x_i} + \gamma_1 \nu^{\pi}(\nu^{\pi}+3) + \gamma_2 \nu^{\nu}(\nu^{\nu}+3) + \delta_1 L_{\pi}(L_{\pi}+1) + \delta_2 L_{\nu}(L_{\nu}+1) + \delta L(L+1) + \varepsilon \Lambda^0_1$$

$$\Lambda^0_1 = \sum_t \frac{1}{2}[c^2_{s;\,t}(\nu^t_s + \frac{1}{2}) + c^2_{d;\,t}(\nu^t + \frac{5}{2})] \tag{21}$$

Similar to procedures which have done to extract the parameters of transitional Hamiltonian in IBM-1 framework, we have supposed $c_d = 1$ and then, Eq. (20) have solved for $i = 1$ case with definite values of $c$ and $\varepsilon$ . Other parameters of Hamiltonian, namely $\delta$ and $\gamma$ , have extracted from empirical available data for *Te* isotopic chain and we would repeat these processes with different values of considered quantities to optimize the $\sigma$ values.

## 2.2. *E0* and *E2* Transition probabilities

The reduced electric monopole and quadrupole transition probabilities are considered as the observables which as well as the quadrupole moment ratios within the low-lying state bands prepare more information about the nuclear structure. The most general single-boson transition operator of angular momentum has the form [5,7]

$$T^{(l)} = \alpha_{2\rho}\delta_{l2}(d^{\dagger}\times s + s^{\dagger}\times d)^{(2)}_{\rho} + \beta_{l\rho}(d^{\dagger}\times d)^{l} + \gamma_{0\rho}\delta_{l0}(s^{\dagger}\times s)^{(0)}_{\rho} \quad , \tag{22}$$

$\rho = \pi$ or $\nu$ which in IBM-2 formalism we consider separate terms for proton and neutron, and $\delta_{l2}$ (and $\delta_{l0}$) are Kronecker deltas. Also $\alpha, \beta$ and $\gamma$ are the constants which are extracted from experimental data and $s^{\dagger}(d^{\dagger})$ represent the creation operator of $s(d)$ boson.. The $T(E0)$ operator may be found by setting $l$=0 in above equation as[5,7]

$$T^{(l)} = \beta_{0\rho}(d^{\dagger}\times d)^{(0)}_{\rho} + \gamma_{0\rho}(s^{\dagger}\times s)^{(0)}_{\rho} \quad , \tag{23}$$

On the other hand, the *E2* transition operator must be a Hermitian tensor of rank two and consequently,

number of bosons must be conserved. With these constraints, there are two operators possible in the lowest order, therefore the electric quadrupole transition operator employed in this study is defined as [7],

$$\hat{T}_{\mu}^{(E2)} = \alpha_{2\rho}\ [\hat{d}^{\dagger} \times \tilde{s}\ +\ \hat{s}^{\dagger} \times \tilde{d}\,]_{\mu}^{(2)} + \beta_{2\rho}\ [\hat{d}^{\dagger} \times \tilde{d}\,]_{\mu}^{(2)} \qquad , \tag{24}$$

To evaluate the $B$(*E0*) and $B$(*E2*) transition ratios and consider the effect of intruder states, we have calculated the matrix elements of $T$(*E0*) and $T$(*E2*) operators between the considered states which are labeled as our model formalism and then, we can extract the constant quantities of Eqs.(23) and (24) in comparison with empirical evidences.

In this study about the considered *Te* isotopes and especially for $B(E2;6_1^+ \to 4_1^+)$, $B(E2;4_1^+ \to 2_1^+)$ and $B(E2;2_1^+ \to 0_1^+)$ transitions which the effects of shape coexistence and intruder states are not dominant, both IBM-1 and IBM-2 methods suggest same accuracy in comparison with experimental counterparts. For excited bands and especially for *E0* transitions which the intruder states are exist, our results show the ability of IBM-2 formalism in the optimization of variation between theoretical predictions and experimental data.

# 3. Theoretical results and comparison with experimental data

## 3.1. Energy levels

The intruder states in *Te* isotopes, especially the $0^+$ states, which are indicated in different analyses where the structure of these nuclei are investigated extensively [5-25], suggest that a mixing of vibrational and rotational, i.e. $U$(5) and $SO$(6) limits in IBM, configurations must be used to explore the experimental energy spectra for these nuclei. This means, the normal vibrational construction would not explore the observed data for two phonon triplet states, $2p-4h$ excitation. Also, the Tellurium isotopes which are located near the proton closed shell are expected to have a spherical shape. In the interacting boson model limits, one can expect, the $U$(5) limits predictions reproduce the energy spectra for such nuclei. The investigations of the experimental energy spectra which have done in Refs.[10-25], suggest the *Te* isotopes as the empirical evidences in transitional region between the spherical, $U$(5) or vibrational, and deformed, $SO$(6) or $\gamma$ -unstable, limits. These results suggest that the normal vibrational states are mixed with deformed intruder states and consequently, one can use a $U(5)\leftrightarrow SO(6)$ transitional Hamiltonians in the determination of energy spectra of this isotopic chain.

To test the disadvantages of using an only spherical prediction for energy spectra of *Te* isotopes, we have used our transitional Hamiltonian, Eq.(16), in the $U(5)$ limit. This means, we have put $c_s = 0$ and then extracted other parameters of Hamiltonian in comparison with experimental data as have explained in previous sections. In Figure1, we have presented the energy spectra of $^{118}Te$ nucleus which have derived by our selected model in comparison with experimental data. A lack of $0_3^+$ state is obvious. Also, our results suggest a very notable variation, $\sigma$ =544, which force us to use a transitional Hamiltonian or in the other words, add the $SO(6)$ terms in Hamiltonian to consider the structure of such nuclei.

In this study, we have used the transitional Hamiltonians which are defined in affine $SU(1,1)$ algebra in IBM-1 and 2 formalisms, *e.g.* Equations 8 and 16 respectively, and then with using the methods which have explained in the previous sections, the parameters of Hamiltonian are extracted in comparison with experimental data. We have considered the 12 levels, $0_1^+$ up to the $2_4^+$ levels for each nucleus determined the $\sigma$ values for these levels. Also, the quantum numbers of these states are taken from Ref.[61]. The best fits for IBM-1 Hamiltonian's parameters, namely $\varepsilon, c_s, \delta$ and $\gamma$ have used in the present work are presented in Table 1. These quantities describe the best agreement between the calculated energy levels in this study and their experimental counterparts which taken from Refs.[66-71], *i.e.* minimum values for $\sigma$ . The control parameter of this model which describes the mixing ratio of spherical and gamma unstable deformed shapes in the structure of nuclei, $c_s$, explore an obvious relation with the variation of our results. The quality of extraction procedures have decreased when the mixing ratio increased and vice versa. When we consider the $SO(6)$ terms in Hamiltonian, $c_s$ will increase from 0 to 1, the deformed intruder states are considered in calculation and therefore, our model suggest more exact results in comparison with experimental counterparts. We have presented a comparison between the results of this study and experimental available data for $^{120}Te$ and $^{122}Te$ isotopes in Figure 2.

In the recent descriptions of intruder states in *Te* isotopes, the IBM-2 predictions have been used to calculate separately the normal and intruder states. To this aim, we have introduced the IBM-2 formalism of transitional Hamiltonian in previous section, *e.g.* Eq.16. We have used a similar procedure as has done for IBM-1 calculations to extract the parameters of transitional Hamiltonian through the experimental counterparts via Bethe-Ansatz method and Least Square (LS) extraction. The quantum numbers of different states of each isotope, the parameters of Hamiltonian and a comparison between theoretical prediction and experimental values for considered isotopes are presented in Tables 2-7, respectively for $^{118-128}Te$ isotopes.

The results of IBM-2 suggest more exact results, i.e. minimum $\sigma$ values, in comparison with experimental data and also the IBM-1 predictions. We have found, the IBM-2 formalism increase the *SO*(6) weight in calculation, *e.g.* $c_s^{IBM-2} > c_s^{IBM-1}$, and therefore, the combination of deformed states with spherical ones optimize our results in comparison with IBM-1. These results may interpret the IBM-2 method as a framework to describe the intruder states in energy spectra of *Te* isotopes with high accuracy. In Figure 3, we have showed the variation of quality factor, $\sigma$, as a function of control parameter in IBM-2 formalism which similar to the results of IBM-1, the best result has occurred at the maximum $c_s$.

We have used a transitional Hamiltonian, in both IBM-1 and IBM-2 versions, to study *Te* isotopes. This formalism has not any inconsistency with other theoretical studies which have done with special hypotheses about mixing of intruder and normal configurations. These $^{118-128}Te$ isotopes which are located near the closed shell have not a significant deformed gamma unstable structure and therefore, our $c_s$ values are set in $c_s$: 0.10 - 0.39 region. This result confirms our theory to use a transitional Hamiltonian and adding the deformed gamma unstable terms in Hamiltonian. Also, the result which are far from the expected values for *SO*(6) limit, $c_s$=1, suggest the dominant effect own to spherical shape and a weak effect due to deformation to describe all normal and intruder states. These results may suggest the $c_s$ value as a measure which describes the mixing of both vibrating and rotating structures in these nuclei and this control parameter has the same role of mixing parameter of other investigation which explain the combination of normal and intruder configurations.

We have tested the relation between the control parameters and the quadrupole, hexadecapole and hexacontatetrapole deformation parameters in this isotopic chain to investigate a relation between shape coexistence and nuclear deformations. In Figures 4(a), (b) and (c), we have showed the variation of our results for control parameters in comparison with quadrupole, hexadecapole and hexacontatetrapole deformation parameters, respectively. These figures describe similar results and suggest an increase in the effect of gamma deformed terms in transitional Hamiltonian, $c_s$ increases from 0 to ~ 0.50 , when the deformation of considered nucleus is increased. These results and especially obvious relation between deformation and shape coexistence may suggest us to consider other versions of IBM such as *sdg*-IBM and *spdg*-IBM in order to describe these hexadecapole and hexacontatetrapole deformations with the meaning of configuration mixing.

Also, we have compared our results with different analyses such as Refs.[10-23] to appreciate the advantages (or disadvantages) of different theoretical methods. These nuclei have not considered in other theoretical studies completely but for some of them which we have found similar counterparts, our results which have derived via more parameters (as compared to the standard IBM shape-phase transition approaches [35-40]) explore more exact results. This means, for these numbers of levels in this energy region, the affine $SU(1,1)$ approach can regard as the more exact method for describing the energy spectra of considered nuclei in transitional region.

### 3.2. $B(E2)$ and $B(E0)$ Transition probabilities

The stable even-even nuclei in *Te* isotopic chain exhibit an excellent opportunity for studying behavior of the total low-lying $E2$ strengths in transitional region from deformed to spherical nuclei. Computation of electromagnetic transition is a sign of good test for nuclear model wave functions. To determine boson effective charges, we have extracted these quantities from the empirical $B(E2)$ values via Least square technique. Extracted values for effective charge parameters have presented in Tables 8 and 9 for IBM-1 and IBM-2 predictions, respectively. For interband transitions between the levels of ground bound which would not contain intruder states, we have used the IBM-1 predictions and as has shown in Figure 5, one can conclude this model describes the elegance of comparison between theoretical prediction and experimental counterparts without the effects of intruder bands.

On the other hand, for excited bounds which we must consider some intruder $0^+$ levels, we have performed our calculations via IBM-2 formalism. Table 11 shows a comparison between our results and experimental counterparts which a satisfactory degree of agreement is achieved. In all Figures, uncertainties of the experimental data which are smaller than the size of symbols are not represented.

From these Figures and Tables, one can conclude, the calculated energy spectra in this approach are generally in good agreements with the experimental data. Our results indicate the elegance of the extraction procedure which has presented in this technique and they suggest the success of estimation processes. Also, theoretical *B*(*E0*) and *B*(*E2*) transition probabilities of even-even *Te* isotopes, which have obtained by using the model perspectives, exhibit nice agreement with experimental ones.

Shape coexistences were associated with sudden changes in the nuclear collective behavior reflected in some quantities such as $R_{4/2} = E(4_1^+) / E(2_1^+)$ *i.e.* rotational excitation. This quantity is expected to be about 2.0 for the spherical nuclei ( the vibrator or *U*(5) limit) and it can increase to 2.50 which is suggested for the deformed $\gamma-$soft or *SO*(6) limit. Similar to what have done in the *Te* isotopes, when $R_{4/2}$ values show deviation from 2.00, the deformation aspects may expect in the structure of these nuclei. Our results which are tabulated in Table 12, suggest some non-zero values for control parameter, namely the ratio of configuration mixing, and also rise when the deformation is increased, approve this meaning. Also, we have considered the vibrational excitation measure, *e.g.* $R_{0/2} = E_{0_2^+} \big/ E_{2_1^+}$ , and two $^{124}Te$ and $^{126}Te$ isotopes have some especial values which are about Iachello's prediction for *E*(5) critical symmetry in the spherical to deformed gamma unstable region [56-57]. On the other hand, these isotopes have the maximum values of control parameters and consequently, the deformed gamma unstable effect has increased and we can expect *E*(5) like features in them.

## 4. Summary and conclusions

We have studied energy spectra and transition probabilities of $^{118\text{-}128}Te$ isotopic chain in the $U(5) \leftrightarrow SO(6)$ transitional region of interacting boson model by a $SU(1,1)$ -based Hamiltonian to describe some intruder states. The validity of the presented parameters in the IBM formulations has been investigated and it is seen that there is an existence of a satisfactory agreement between presented results and experimental counterparts. We may conclude that general characteristics of *Te* isotopes are well accounted in this study and the idea of shape coexistence in this region is supported. Our results suggest that gamma-soft rotor features exist in *Te* isotopes, but with a dominancy of vibrational character where $^{124}Te$ and $^{124}Te$ show $E(5)$ like structures. The results of this study confirm that this technique is worth extending for investigating the nuclear structure of other nuclei existing around the mass of $A \sim 130$.

## Acknowledgement

This work is published as a part of research project supported by the University of Tabriz Research Affairs Office.

## Tables

Table1. The parameters of IBM-1 Hamiltonian for different $Te$ isotopes which are extracted by least square method from experimental data. $N$ describes the boson number and. $\sigma$ is regarded as the quality for extraction processes.

| Nucleus | $N$ | $\varepsilon(kev)$ | $c_s$ | $\gamma(kev)$ | $\delta(kev)$ | $\sigma$ |
|---|---|---|---|---|---|---|
| $^{118}_{52}Te$ | 9 | 840 | 0.03 | –27.33 | 9.44 | 309 |
| $^{120}_{52}Te$ | 8 | 880 | 0.12 | –31.06 | 10.65 | 273 |
| $^{122}_{52}Te$ | 7 | 822 | 0.15 | –19.50 | 6.87 | 268 |
| $^{124}_{52}Te$ | 6 | 800 | 0.28 | –20.88 | 12.33 | 173 |
| $^{126}_{52}Te$ | 5 | 820 | 0.19 | –15.22 | 14.20 | 233 |
| $^{128}_{52}Te$ | 4 | 830 | 0.17 | –20.97 | 8.55 | 259 |

Table2. Energy spectra of $^{118}Te$ nucleus which are determined by transitional Hamiltonian and the experimental values are taken from Refs.[65-66]. Parameters of IBM-2 Hamiltonian in this case are $N$ = 9, $\varepsilon$ = 780 *kev*, $\gamma_1$= -15.44 *kev*, $\gamma_2$=-17.76 *kev* , $\delta_1$=5.24 *kev*, $\delta_2$=11.80 *kev* and $\delta$=2.97 *kev*.

| level | $k$ | $v^{\pi}$ | $v^{\nu}$ | $L_{\pi}$ | $L_{\nu}$ | $E_{\mathrm{exp}}(kev)$ | $E_{\mathrm{th}}(kev)$ |
|---|---|---|---|---|---|---|---|
| $0_1^+$ | 4 | 0 | 0 | 0 | 0 | 0 | 0 |
| $2_1^+$ | 3 | 1 | 1 | 1 | 2 | 606 | 687 |
| $4_1^+$ | 3 | 2 | 1 | 2 | 2 | 1206 | 1271 |
| $6_1^+$ | 3 | 1 | 3 | 2 | 3 | 1821 | 1896 |
| $3_1^+$ | 3 | 3 | 1 | 3 | 1 | 1892 | 1948 |
| $0_2^+$ | 4 | 0 | 1 | 0 | 0 | 957 | 1009 |
| $2_2^+$ | 3 | 1 | 2 | 1 | 2 | 1151 | 1214 |
| $4_2^+$ | 3 | 3 | 1 | 2 | 2 | 1702 | 1729 |
| $0_3^+$ | 3 | 1 | 3 | 0 | 3 | 1517 | 1608 |
| $2_3^+$ | 4 | 1 | 1 | 2 | 1 | 1482 | 1497 |
| $0_4^+$ | 4 | 0 | 2 | 0 | 0 | - | 1995 |
| $2_4^+$ | 3 | 2 | 1 | 2 | 1 | - | 2074 |
| | | $\sigma$ = 181 | | | $c_s$= 0.10 | | |

Table3. Energy spectra of $^{120}Te$ nucleus which are determined by transitional Hamiltonian and the experimental values are taken from Ref.[67]. Parameters of IBM-2 Hamiltonian in this case are $N$ = 8, $\varepsilon$ = 755 *kev*, $\gamma_1$ = -11.37 *kev*, $\gamma_2$ = -16.42 *kev* , $\delta_1$= 9.20 *kev*, $\delta_2$ = 10.58 *kev* and $\delta$ = 8.09 *kev*.

| level | $k$ | $v^{\pi}$ | $v^{\nu}$ | $L_{\pi}$ | $L_{\nu}$ | $E_{\mathrm{exp}}(kev)$ | $E_{\mathrm{th}}(kev)$ |
|---|---|---|---|---|---|---|---|
| $0_1^+$ | 4 | 0 | 0 | 0 | 0 | 0 | 0 |
| $2_1^+$ | 3 | 1 | 1 | 1 | 2 | 560 | 615 |
| $4_1^+$ | 3 | 2 | 1 | 2 | 2 | 1161 | 1240 |
| $6_1^+$ | 2 | 3 | 1 | 3 | 2 | 1776 | 1855 |
| $3_1^+$ | 2 | 3 | 1 | 3 | 1 | 1863 | 2002 |
| $0_2^+$ | 4 | 0 | 1 | 0 | 0 | 1103 | 1266 |
| $2_2^+$ | 3 | 2 | 1 | 1 | 2 | 1201 | 1331 |
| $4_2^+$ | 2 | 3 | 1 | 2 | 2 | 1815 | 1899 |
| $0_3^+$ | 2 | 1 | 3 | 0 | 3 | 1613 | 1675 |
| $2_3^+$ | 4 | 1 | 1 | 2 | 1 | 1535 | 1711 |
| $0_4^+$ | 4 | 0 | 2 | 0 | 0 | - | 1863 |
| $2_4^+$ | 3 | 2 | 1 | 2 | 1 | 1924 | 2118 |

$\sigma$ = 139 $c_s$= 0.21

Table4. Energy spectra of $^{122}Te$ nucleus which are determined by transitional Hamiltonian and the experimental values are taken from Refs.[65-68]. Parameters of IBM-2 Hamiltonian in this case are $N$ = 7, $\varepsilon$ = 800 *kev*, $\gamma_1$ = - 7.28 *kev*, $\gamma_2$ = -11.43 *kev* , $\delta_1$ = 9.66 *kev*, $\delta_2$ =7.05 *kev* and $\delta$ =10.33 *kev*.

| level | $k$ | $v^{\pi}$ | $v^{\nu}$ | $L_{\pi}$ | $L_{\nu}$ | $E_{\mathrm{exp}}(kev)$ | $E_{\mathrm{th}}(kev)$ |
|---|---|---|---|---|---|---|---|
| $0_1^+$ | 3 | 0 | 0 | 0 | 0 | 0 | 0 |
| $2_1^+$ | 3 | 1 | 1 | 1 | 2 | 564 | 611 |
| $4_1^+$ | 2 | 2 | 1 | 2 | 2 | 1181 | 1255 |
| $6_1^+$ | 2 | 1 | 3 | 2 | 3 | 1751 | 1847 |
| $3_1^+$ | 2 | 3 | 1 | 3 | 1 | 1952 | 2095 |
| $0_2^+$ | 3 | 0 | 1 | 0 | 0 | 1357 | 1394 |
| $2_2^+$ | 2 | 1 | 2 | 1 | 2 | 1256 | 1387 |
| $4_2^+$ | 2 | 3 | 1 | 2 | 2 | 1910 | 2003 |
| $0_3^+$ | 2 | 1 | 3 | 0 | 0 | 1747 | 1842 |
| $2_3^+$ | 3 | 1 | 1 | 2 | 1 | 1752 | 1905 |
| $0_4^+$ | 3 | 0 | 2 | 0 | 0 | 1940 | 2063 |
| $2_4^+$ | 2 | 2 | 1 | 2 | 1 | 2099 | 2210 |

$\sigma$ = 114 $c_s$= 0.26

Table5. Energy spectra of $^{124}Te$ nucleus which are determined by transitional Hamiltonian and the experimental values are taken from Refs.[65-69]. Parameters of IBM-2 Hamiltonian in this case are $N = 6$, $\varepsilon = 800$ *kev*, $\gamma_1 = -20.01$ *kev*, $\gamma_2 = -9.88$ *kev* , $\delta_1 = 10.04$ *kev*, $\delta_2 = 19.83$ *kev* and $\delta = 6.11$ *kev*.

| level | $k$ | $v^{\pi}$ | $v^{\nu}$ | $L_{\pi}$ | $L_{\nu}$ | $E_{\text{exp}}(kev)$ | $E_{\text{th}}(kev)$ |
|---|---|---|---|---|---|---|---|
| $0_1^+$ | 3 | 0 | 0 | 0 | 0 | 0 | 0 |
| $2_1^+$ | 3 | 1 | 1 | 1 | 2 | 603 | 682 |
| $4_1^+$ | 2 | 1 | 2 | 2 | 2 | 1249 | 1207 |
| $6_1^+$ | 2 | 1 | 3 | 2 | 3 | 1747 | 1901 |
| $3_1^+$ | 2 | 3 | 1 | 3 | 1 | 2039 | 2115 |
| $0_2^+$ | 3 | 0 | 1 | 0 | 0 | 1657 | 1709 |
| $2_2^+$ | 2 | 1 | 2 | 1 | 2 | 1326 | 1398 |
| $4_2^+$ | 2 | 3 | 1 | 2 | 2 | 1958 | 2013 |
| $0_3^+$ | 2 | 1 | 3 | 0 | 0 | 1883 | 1967 |
| $2_3^+$ | 3 | 1 | 1 | 2 | 1 | 2039 | 2086 |
| $0_4^+$ | 3 | 0 | 2 | 0 | 0 | 2153 | 2249 |
| $2_4^+$ | 2 | 2 | 1 | 2 | 1 | 2182 | 2258 |
| | | | $\sigma = 83$ | | | $c_s = 0.39$ | |

Table 6. Energy spectra of $^{126}Te$ nucleus which are determined by transitional Hamiltonian and the experimental values are taken from Ref.[70]. Parameters of IBM-2 Hamiltonian in this case are $N = 5$, $\varepsilon = 865$ *kev*, $\gamma_1 = -6.13$*kev*, $\gamma_2 = -14.35$ *kev* , $\delta_1 = 10.22$ *kev*, $\delta_2 = 9.07$ *kev* and $\delta = 4.77$*kev*.

| level | $k$ | $v^{\pi}$ | $v^{\nu}$ | $L_{\pi}$ | $L_{\nu}$ | $E_{\text{exp}}(kev)$ | $E_{\text{th}}(kev)$ |
|---|---|---|---|---|---|---|---|
| $0_1^+$ | 2 | 0 | 0 | 0 | 0 | 0 | 0 |
| $2_1^+$ | 2 | 1 | 1 | 1 | 2 | 743 | 812 |
| $4_1^+$ | 1 | 2 | 1 | 2 | 2 | 1497 | 1582 |
| $6_1^+$ | 1 | 1 | 3 | 2 | 3 | 1811 | 1900 |
| $3_1^+$ | 1 | 3 | 1 | 3 | 1 | 2164 | 2256 |
| $0_2^+$ | 2 | 0 | 1 | 0 | 0 | 1978 | 2055 |
| $2_2^+$ | 1 | 1 | 2 | 1 | 2 | 1520 | 1569 |
| $4_2^+$ | 1 | 3 | 1 | 2 | 2 | 2028 | 2166 |
| $0_3^+$ | 3 | 1 | 3 | 0 | 3 | - | 1493 |
| $2_3^+$ | 2 | 1 | 1 | 2 | 1 | 2193 | 2251 |
| $0_4^+$ | 2 | 0 | 2 | 0 | 0 | - | 1866 |
| $2_4^+$ | 3 | 2 | 1 | 2 | 1 | 2218 | 2305 |
| | | | $\sigma = 103$ | | | $c_s = 0.35$ | |

Table7. Energy spectra of $^{128}Te$ nucleus which are determined by transitional Hamiltonian and the experimental values are taken from Refs.[65-71]. Parameters of IBM-2 Hamiltonian in this case are $N = 4$, $\varepsilon = 875$ *kev*, $\gamma_1 = -9.33$ *kev*, $\gamma_2 = -7.66$ *kev* , $\delta_1 = 8.22$ *kev*, $\delta_2 = 16.52$ *kev* and $\delta = 1.75$*kev*.

| level | $k$ | $v^{\pi}$ | $v^{\nu}$ | $L_\pi$ | $L_\nu$ | $E_{\exp}(kev)$ | $E_{\mathrm{th}}(kev)$ |
|---|---|---|---|---|---|---|---|
| $0_1^+$ | 2 | 0 | 0 | 0 | 0 | 0 | 0 |
| $2_1^+$ | 2 | 1 | 1 | 1 | 2 | 839 | 911 |
| $4_1^+$ | 1 | 2 | 1 | 2 | 2 | 1633 | 1729 |
| $6_1^+$ | 1 | 1 | 3 | 2 | 3 | 1815 | 1874 |
| $3_1^+$ | 1 | 3 | 1 | 3 | 1 | 2139 | 2197 |
| $0_2^+$ | 2 | 0 | 1 | 0 | 0 | 1965 | 2014 |
| $2_2^+$ | 1 | 1 | 2 | 1 | 2 | 1583 | 1663 |
| $4_2^+$ | 1 | 3 | 1 | 2 | 2 | 1982 | 2077 |
| $0_3^+$ | 3 | 1 | 3 | 0 | 3 | - | 1608 |
| $2_3^+$ | 2 | 1 | 1 | 2 | 1 | 1886 | 1983 |
| $0_4^+$ | 2 | 0 | 2 | 0 | 0 | - | 2007 |
| $2_4^+$ | 3 | 2 | 1 | 2 | 1 | 2190 | 2278 |
| | | | $\sigma = 109$ | | $c_s = 0.29$ | | |

Table9. The effective quadrupole coefficients of considered isotope which are extracted from experimental data by Eq.(24) in IBM-1 framework.

| Nucleus | $^{118}Te$ | $^{120}Te$ | $^{122}Te$ | $^{124}Te$ | $^{126}Te$ | $^{128}Te$ |
|---|---|---|---|---|---|---|
| $q_2$ | 1.883 | 2.013 | 2.217 | 2.338 | 2.305 | 2.261 |
| $q_2'$ | –0.927 | –0.858 | –0.801 | –0.709 | –0.766 | –0.794 |

Table10. The effective quadrupole coefficients of considered isotope which are extracted from experimental data by Eq.(24) in IBM-2 framework.

| Nucleus | $^{118}Te$ | $^{120}Te$ | $^{122}Te$ | $^{124}Te$ | $^{126}Te$ | $^{128}Te$ |
|---|---|---|---|---|---|---|
| $e_\pi$ | 2.366 | 2.305 | 2.113 | 1.755 | 1.934 | 2.105 |
| $e_\nu$ | 0.922 | 0.849 | 0.794 | 0.486 | 0.602 | 0.715 |
| $q_2$ | 1.289 | 1.461 | 1.599 | 2.115 | 2.009 | 1.778 |
| $q_2'$ | –1.703 | –1.535 | –1.457 | –0.968 | –1.098 | –1.207 |

Table11. IBM-2 predictions for *B*(*E0*) and *B*(*E2*) transition probabilities are presented for some transitions between excited bounds which their experimental counterparts are available. Experimental values are taken form Refs.[5,7,65-72].

| Nucleus | Transition probability | Exp. | The. |
|---|---|---|---|
| $^{118}Te$ | $B(E0;\ 0_2^+ \to 0_1^+)$ | 0.40 | 0.42 |
| | $B(E0;\ 0_3^+ \to 0_2^+)$ | 0.031 | 0.037 |
| | $B(E2;\ 2_2^+ \to 0_2^+)$ | 0.31 | 0.36 |
| $^{120}Te$ | $B(E0;\ 0_2^+ \to 0_1^+)$ | 0.103 | 0.110 |
| | $B(E0;\ 0_3^+ \to 0_2^+)$ | 1.50 | 1.55 |
| | $B(E0;\ 2_2^+ \to 2_1^+)$ | 0.12 | 0.18 |
| | $B(E2;\ 2_2^+ \to 0_1^+)$ | 0.15 | 0.21 |
| | $B(E2;\ 2_2^+ \to 0_2^+)$ | 0.50 | 0.58 |
| $^{122}Te$ | $B(E0;\ 0_2^+ \to 0_1^+)$ | 0.139 | 0.147 |
| | $B(E0;\ 0_3^+ \to 0_2^+)$ | 0.132 | 0.139 |
| | $B(E0;\ 0_4^+ \to 0_3^+)$ | 0.013 | 0.011 |
| | $B(E2;\ 2_2^+ \to 0_1^+)$ | 0.841 | 0.853 |
| | $B(E2;4_2^+ \to 2_2^+)$ | 0.61 | 0.67 |
| | $B(E2;\ 2_3^+ \to 2_2^+)$ | 0.21 | 0.27 |
| $^{124}Te$ | $B(E0;\ 0_2^+ \to 0_1^+)$ | 0.11 | 0.14 |
| | $B(E0;\ 0_3^+ \to 0_2^+)$ | 0.001 | 0.002 |
| | $B(E2;\ 2_2^+ \to 2_1^+)$ | 0.019 | 0.021 |
| $^{126}Te$ | $B(E0;\ 2_2^+ \to 2_1^+)$ | 0.52 | 0.55 |

Table12. IBM-2 predictions for control parameter, rotational and vibrational excitations are presented for considered nuclei. These results suggest the $^{124}Te$ and $^{126}Te$ isotopes as nuclei which the deformation effect is increased in their structures.

| Nucleus | $^{118}Te$ | $^{120}Te$ | $^{122}Te$ | $^{124}Te$ | $^{126}Te$ | $^{128}Te$ |
|---|---|---|---|---|---|---|
| $c_s$ | 0.10 | 0.21 | 0.26 | 0.39 | 0.35 | 0.29 |
| $R_{0/2} = E_{0_2^+} / E_{2_1^+}$ (vibrational excitation) | 1.62 | 1.93 | 2.49 | 2.86 | 2.83 | 2.66 |
| $R_{4/2} = E_{4_1^+} / E_{2_1^+}$ (Rotational excitation) | 1.99 | 2.07 | 2.09 | 2.07 | 2.04 | 2.01 |

## Figure caption

Figure1. Comparison between the predictions of $U(5)$ model for energy spectra of $^{118}Te$ and experimental counterparts were taken from Refs.[65-66]. Due to the similar results for other nuclei, we do not present their spectra. The lack of $0_3^+$ is obvious in theoretical predictions.

Figure2. Theoretical predictions based on IBM-1 results of transitional Hamiltonian for energy levels of $^{120\text{-}122}Te$ isotopes in comparison with experimental data.

Figure3. Variation of quality factor, $\sigma$, is presented for this isotopic chain as a function of control parameters. The IBM-2 predictions suggest the most accuracy for $^{124}Te$ which has $c_s$=0.39.

Figure 4 (contour plots). We have presented the variation of control parameters and quality factors of extraction processes as a function of (a) quadrupole, (b) hexadecapole and (c) hexacontatetrapole deformations for this isotopic chain. All Figures show similar results and suggest more exact results, minimum $\sigma$ values, for systems with high deformation and maximum values of control parameters.

Figure5 (color online). A comparison between the predictions of IBM-1 for $B(E2)$ transition probabilities between ground band levels and experimental counterparts which are taken from Refs.[5,7,65-72].

Figure1.

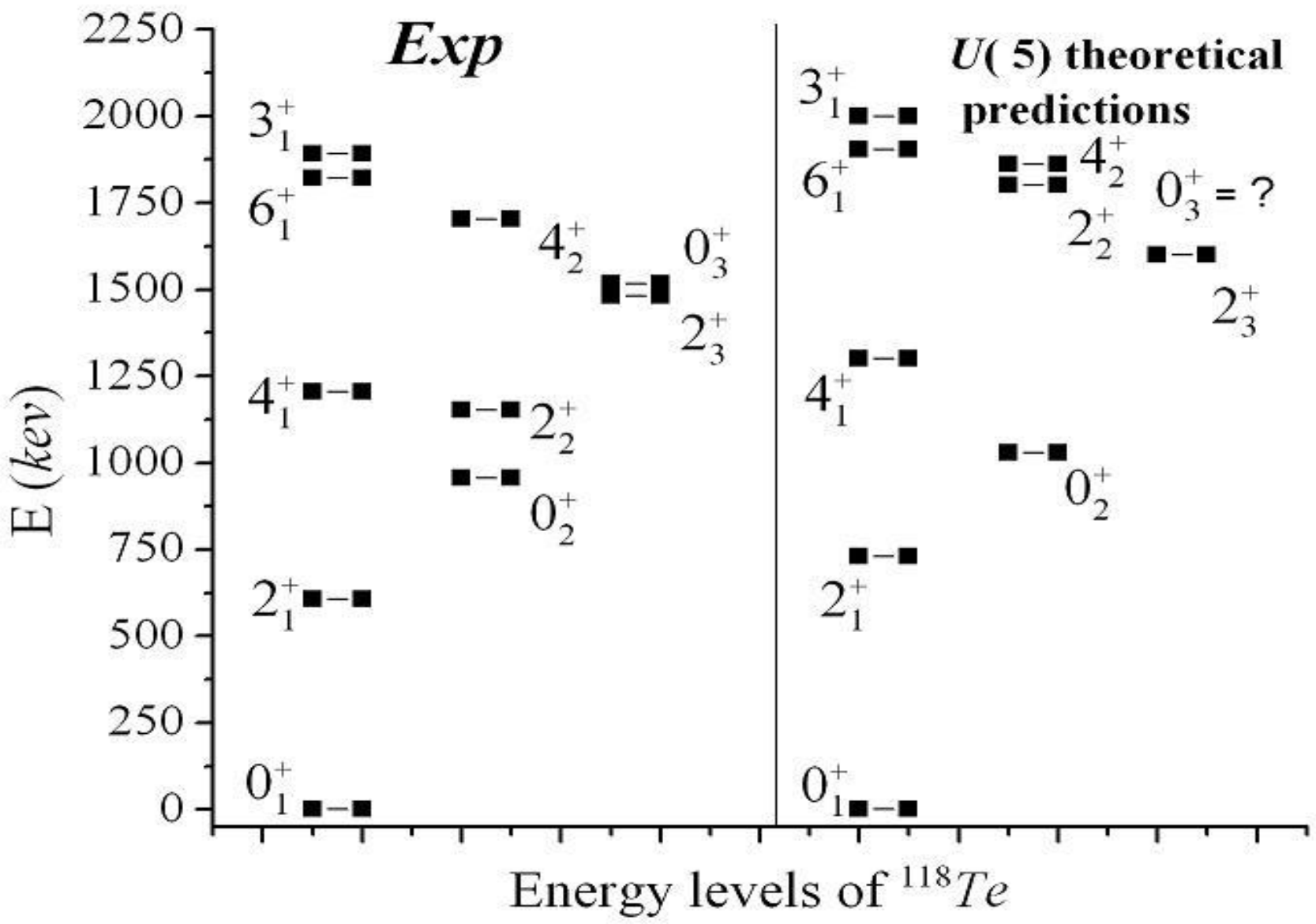

Figure 2.

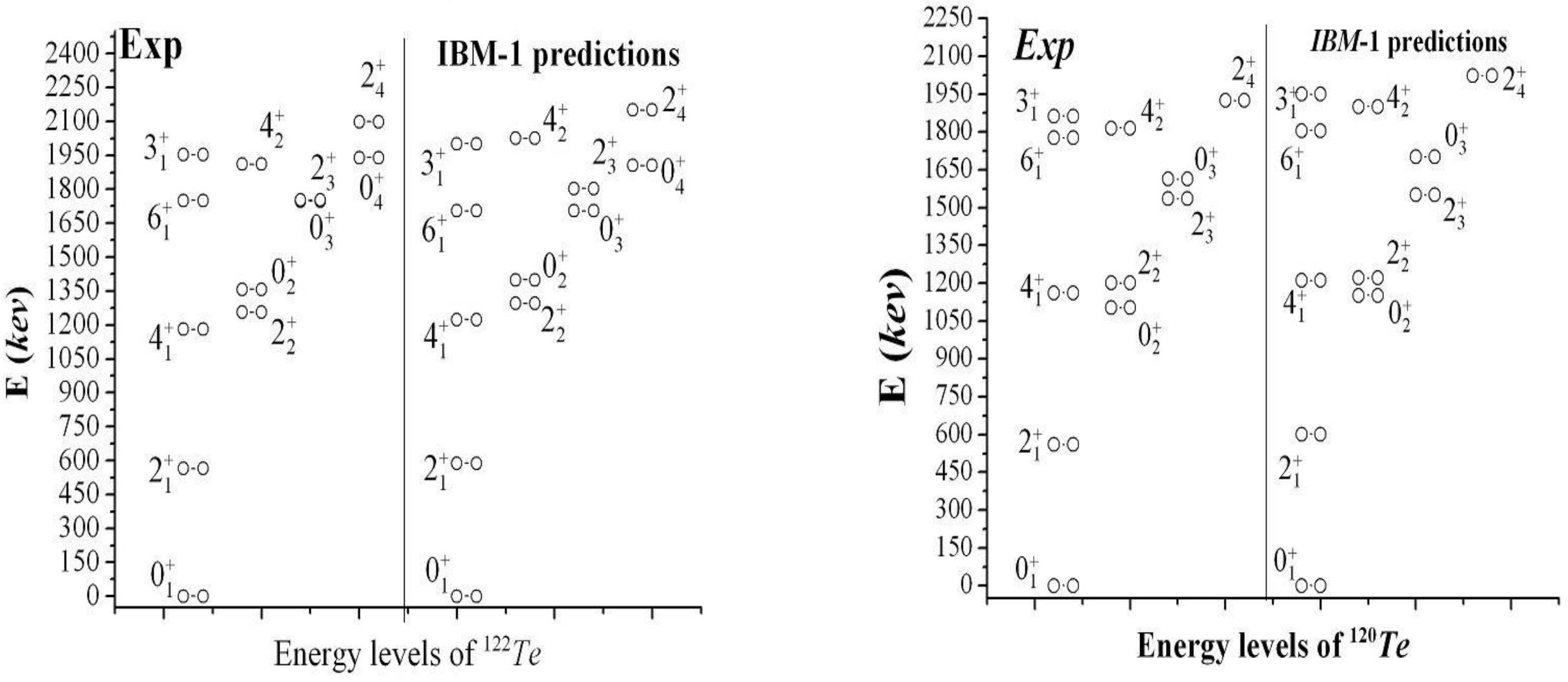

Figure3.

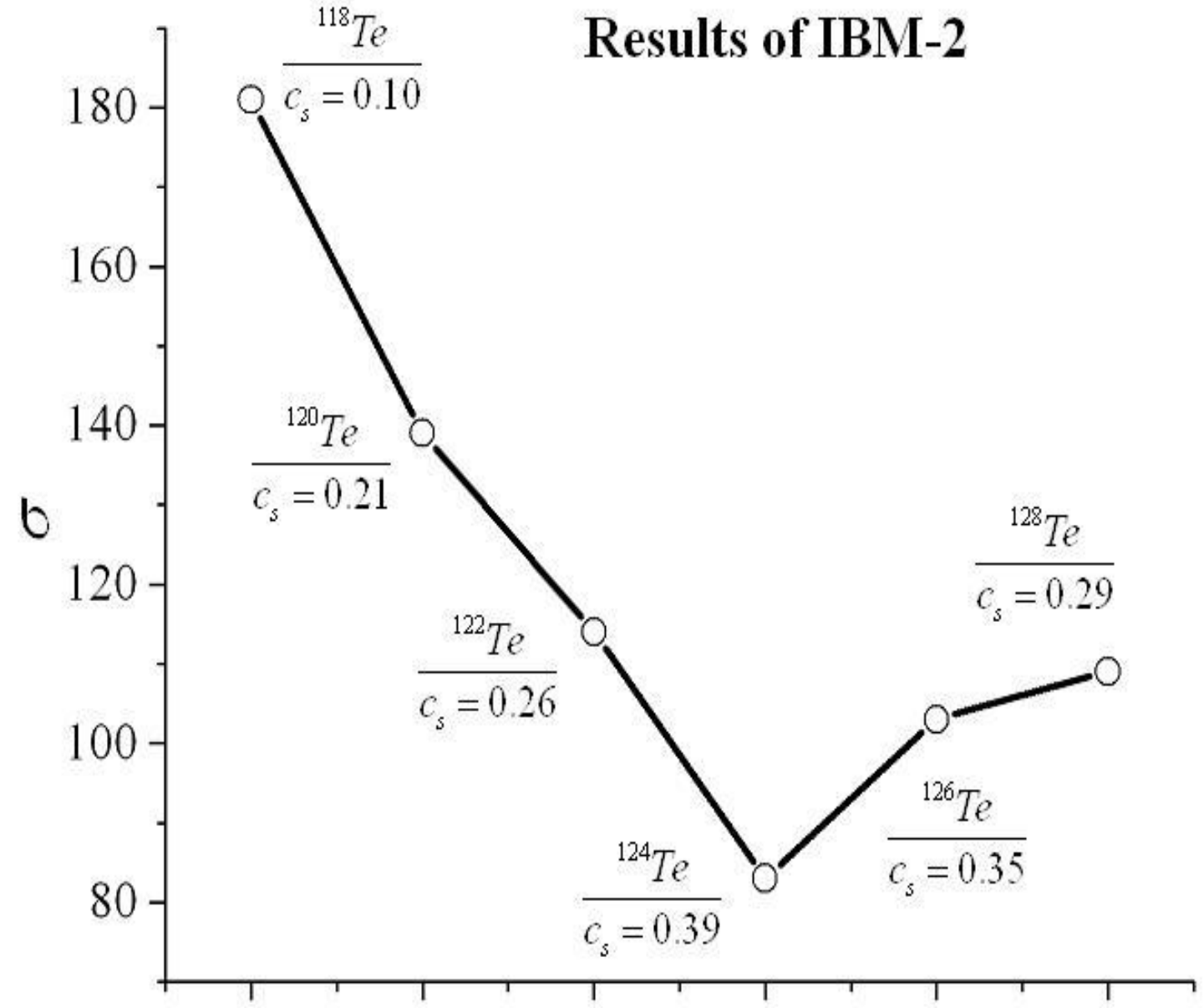

Figure4 (a).

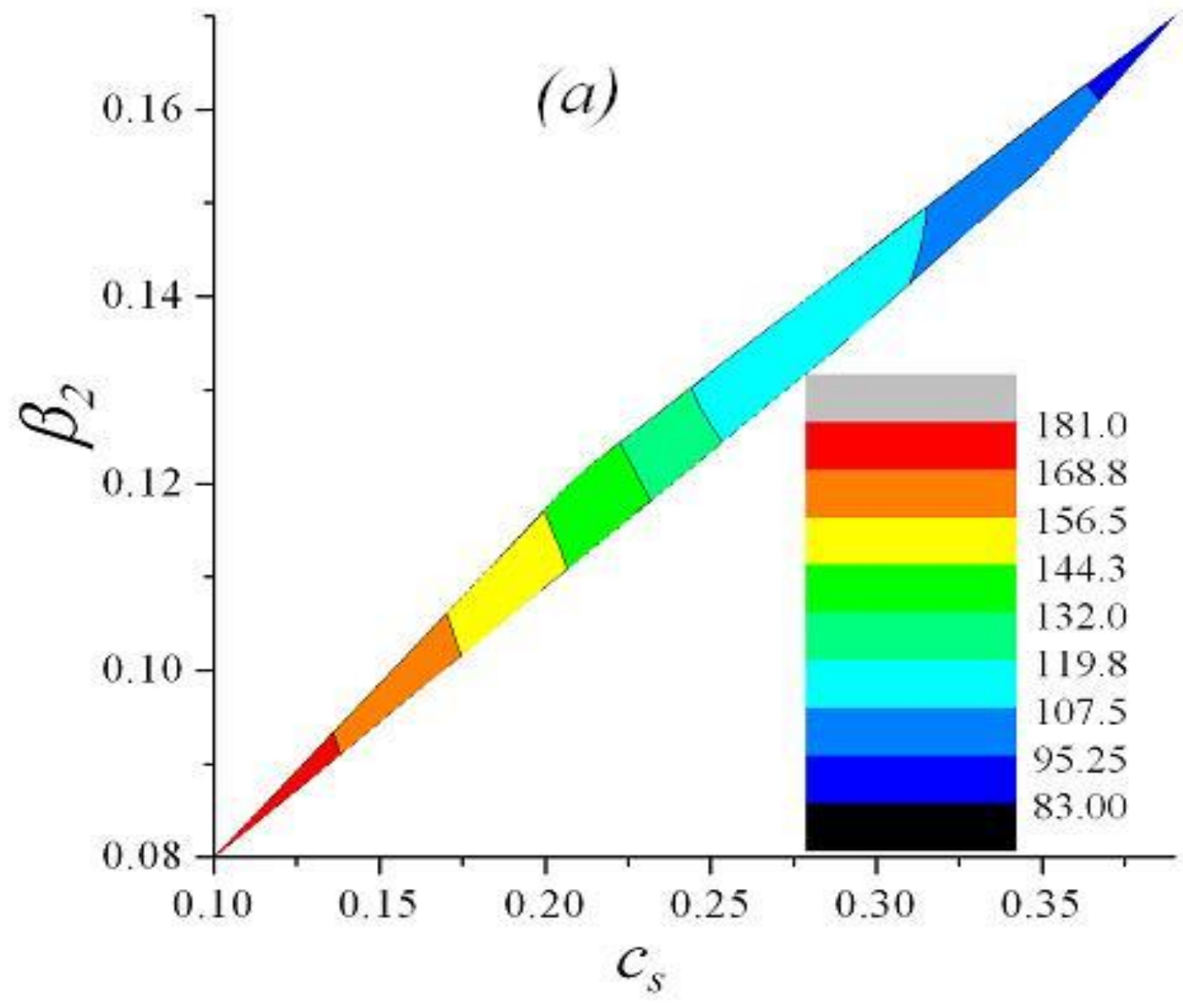

Figure4(b).

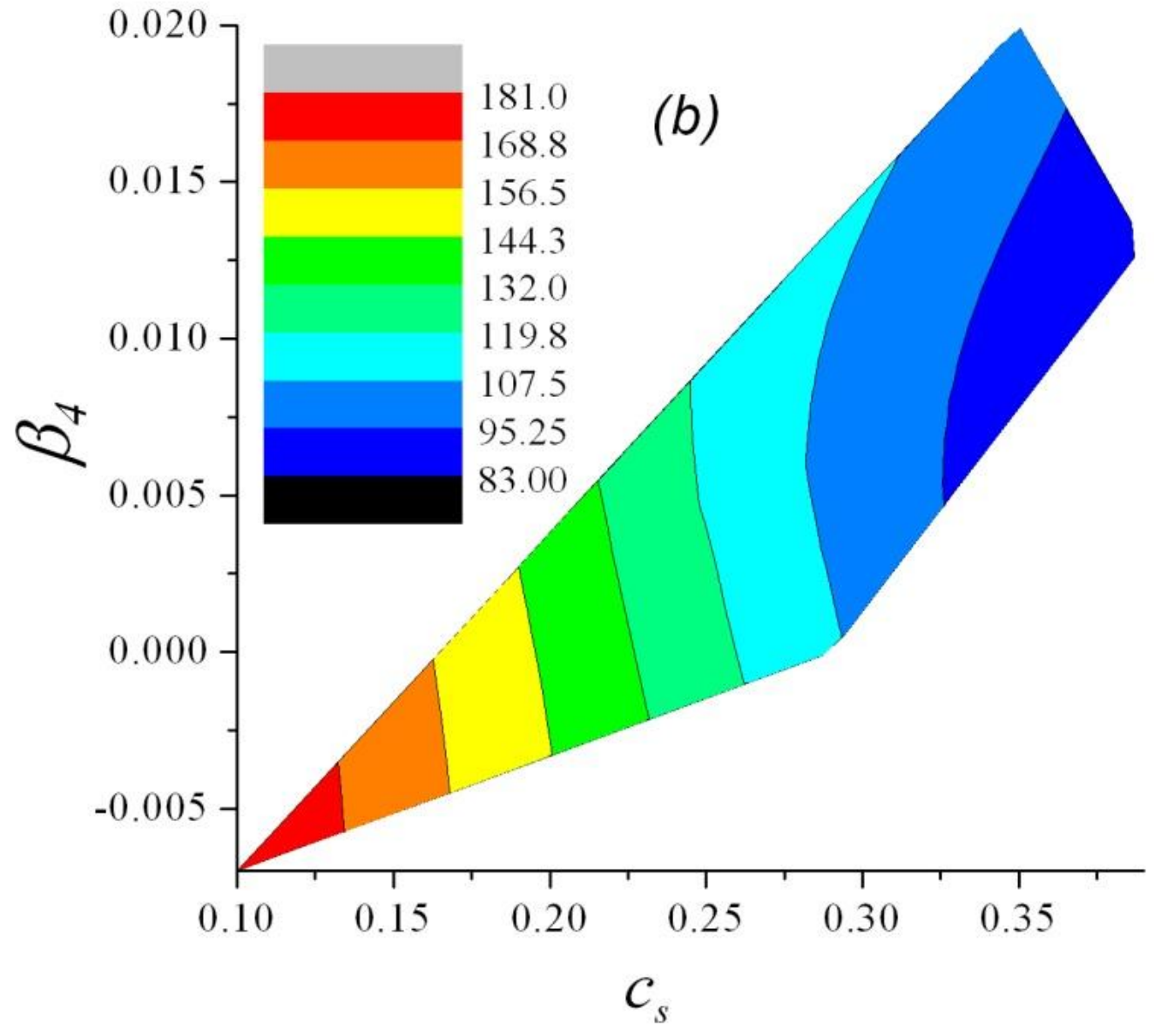

Figure 4(c).

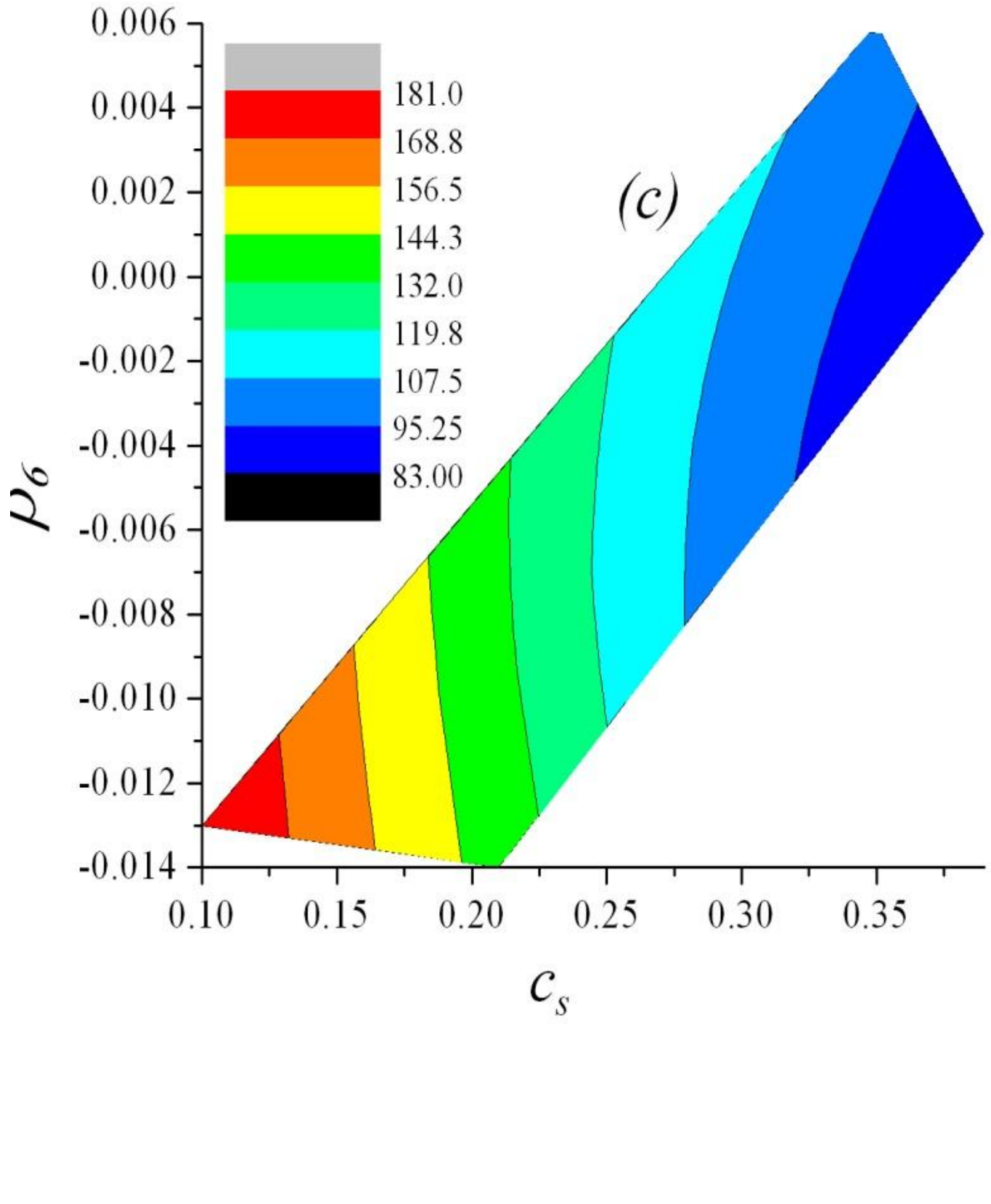

Figure5.

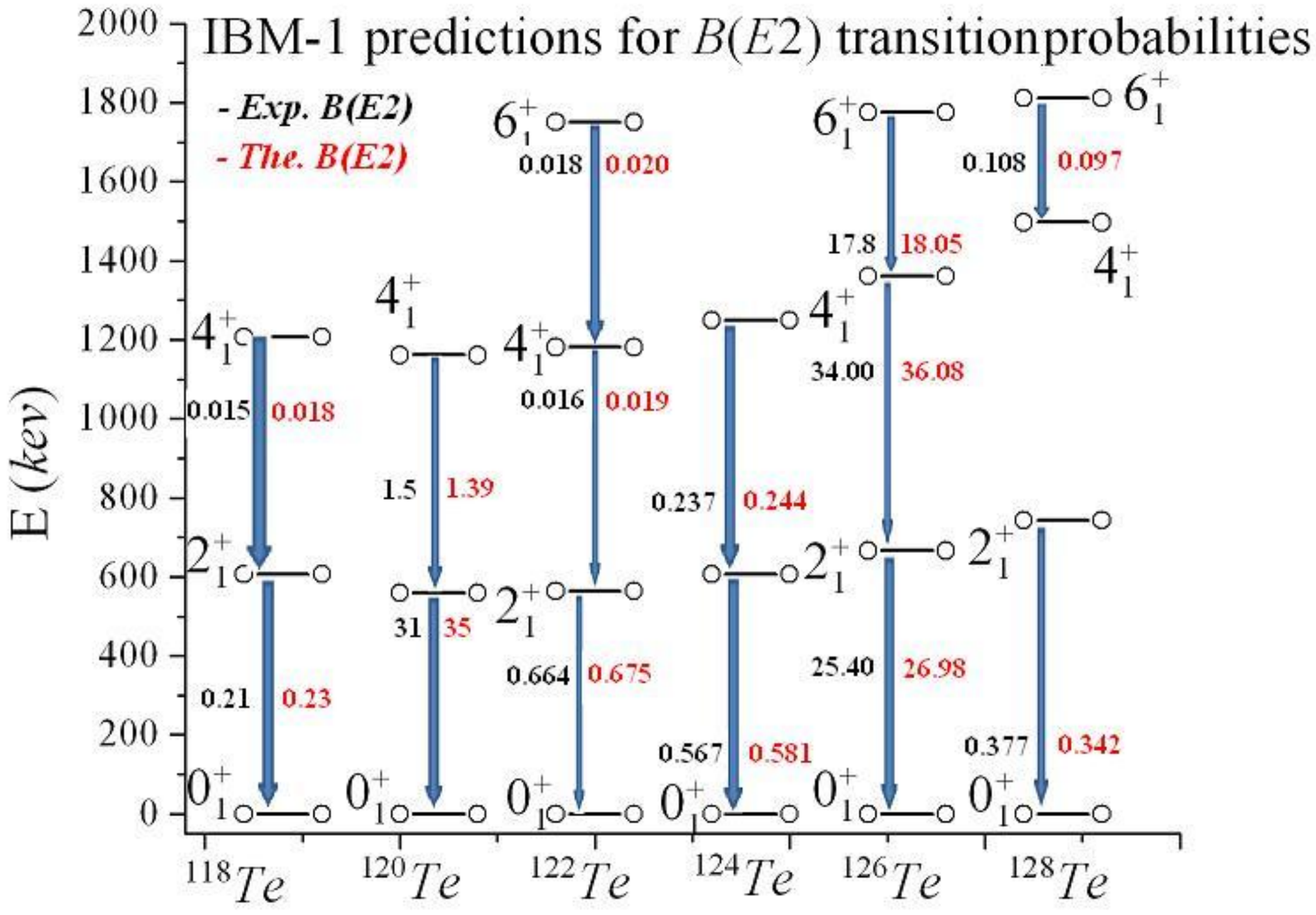